\documentclass[usenatbib]{mn2e}
\usepackage{graphicx,amsmath,multirow,amssymb}
\usepackage{subfig}
\usepackage{natbib}
\newcommand{\comment}[1]{}

\def\simgt{\lower.5ex\hbox{$\; \buildrel > \over \sim \;$}}
\def\simlt{\lower.5ex\hbox{$\; \buildrel < \over \sim \;$}}

\newcommand{\Msun}{\ensuremath{{\rm M}_{\sun}}}

\title[PNe in the SMC]{Planetary Nebulae in the Small Magellanic Cloud}
\author[Ventura et al.]{P. Ventura$^1$, L. Stanghellini$^{2}$, M. Di Criscienzo$^1$,
D. A. Garc\'{\i}a--Hern\'andez$^{3,4}$, 
\newauthor
F. Dell'Agli$^{1}$   \\ \\
$^1$INAF -- Osservatorio Astronomico di Roma, Via Frascati 33, 00040, Monte Porzio Catone (RM), Italy \\
$^2$National Optical Astronomy Observatory, 950 N. Cherry Avenue, Tucson (AZ) 85719, USA \\
$^{3}$Instituto de Astrof\'{\i}sica de Canarias, E-38205 La Laguna, Tenerife, Spain \\
$^{4}$Departamento de Astrof\'{\i}sica, Universidad de La Laguna (ULL), E-38206 La Laguna, Tenerife, Spain\\
}

\begin{document}

\date{Accepted, Received; in original form }

\pagerange{\pageref{firstpage}--\pageref{lastpage}} \pubyear{2012}

\maketitle

\label{firstpage}

\begin{abstract}
We analyse the planetary nebulae (PNe) population of the Small Magellanic Cloud (SMC), 
based on evolutionary models of stars with metallicities in the range 
$10^{-3} \leq Z \leq 4\times 10^{-3}$ and mass $0.9~\Msun < M < 8~\Msun$, 
evolved through the asymptotic giant branch (AGB) phase. The models used account for 
dust formation in the circumstellar envelope. To characterise the PNe sample of the SMC, 
we compare the observed abundances of the various species with the final chemical 
composition of the AGB models: this study allows us to identify the progenitors of the
PNe observed, in terms of mass and chemical composition. According to our interpretation,
most of the PNe descend from low-mass ($M < 2 \Msun$) stars, which become carbon rich, 
after experiencing repeated third dredge-up episodes, during the AGB phase. A fraction
of the PNe showing the signature of advanced CNO processing are interpreted as the progeny 
of massive AGB stars, with mass above $\sim 6 \Msun$, undergoing strong hot bottom burning. 
The differences 
with the chemical composition of the PNe population of the Large Magellanic Cloud (LMC)
is explained on the basis of the diverse star
formation history and age-metallicity relation of the two galaxies. The implications of
the present study for some still highly debated points regarding the AGB evolution are
also commented.
\end{abstract}

140.252.118.146

\section{Introduction}
The final phases in the evolution of stars of intermediate mass 
($0.8\Msun \leq M \leq 8\Msun$) is still poorly known, owing to the uncertainties 
affecting the understanding of the AGB phase. The two main factors preventing reliable AGB modeling 
are the treatment of the convective instability, and that of the mass loss mechanism, 
which are still rather uncertain \citep{herwig05, karakas14}. As a consequence, the role 
played by AGB stars in several astrophysical contexts is still highly debated. On the other 
hand, understanding the AGB impact is crucial for several astrophysical fields, such as the 
determination of the masses of galaxies at high 
redshifts \citep{maraston06}, the formation and chemical evolution of galaxies 
\citep{romano10, santini14}, the dust content of high-redshift quasars \citep{valiante11}, 
and the formation of multiple populations of stars in globular clusters \citep{ventura01}.

Following the pioneering studies by the Heidelberg group \citep{fg06}, some Authors
have recently made significant progresses in AGB modelling when including  the formation 
of dust in the circumstellar envelopes of AGB stars
\citep{paperI, paperII, paperIII, paperIV, nanni13a,nanni13b,nanni14}.

Testing these models requires comparison with the observations. The Magellanic
Clouds (MC) have so far offered the best comparison laboratory, owing to their relatively 
short distances (51 kpc and 61 kpc respectively, for the Large and Small Magellanic Cloud, 
Cioni et al. 2000; Keller \& Wood 2006) and the low reddening ($E_{B-V}=0.15$~mag and 0.04 
mag, respectively, for the LMC and SMC, Westerlund 1997). Furthermore, the study of the MC
populations extends to a wide range of metallicities, broadening the possible comparison 
between data and models, and are less affected by interstellar extinction than Galactic 
populations.

The near- and mid-infrared observations of AGB stars in the MC, compared with the
evolutionary models produced by various research teams, allowed the empirical 
calibration of the physical mechanisms relevant to the evolution of these objects. For example, 
such comparisons have allowed primarily to estimate the extension of the third dredge-up 
(hereinafter TDU), i.e. the inwards penetration of the bottom of the convective envelope 
in the phase following each thermal pulse \citep{izzard04, girardi07, martin07, riebel10, 
riebel12, srinivasan09, srinivasan11, boyer11, boyer12}. More recent studies were focused 
on the dust production by AGB stars in the MC and the characterisation of the most obscured 
sources \citep{flavia14, flavia15a, flavia15b, ventura15, ventura16}.

The study of planetary nebulae offers an alternative opportunity to 
constrain the evolutionary models that evolve through the AGB. In fact, the surface 
chemistry of AGB stars will eventually contribute to the chemical mix of 
the ejected PNe.  By studying the PN abundances we can infer the combined effects of the 
physical mechanisms potentially able to alter the surface chemical composition of AGB stars. 
In particular, two main processes are active in this chemical transformation
of the stellar surface, namely, the TDU, and the hot bottom burning (HBB), i.e.
the proton capture nucleosynthesis active at the base of the envelope of the stars
whose initial mass is above $\sim 4~\Msun$ \citep{renzini81, blocker91}. Knowledge of
the nebular chemical composition of PNe allows to understand the relative importance of 
these phenomena on the modification of the surface chemical composition during the whole
AGB evolution \citep{marigo03, marigo11}. Another advantage of using PNe to constrain 
AGB evolutionary models resides in the fact that PN abundance analysis is 
relatively straightforward in comparison to that of AGB stars, whose
optical and near-IR spectra are contaminated by millions of molecular lines, which render
extremely complex the derivation of the abundances of individual species 
\citep[e.g.][]{garcia06, garcia07, garcia09}.

Since the evolutionary properties of AGB stars might depend on the environment
where they have formed and evolved (e.g. the initial medium composition, the density of 
stars, etc), it is mandatory to study the nebular abundances in different PN populations 
(e.g., at several metallicities) and compare them with models of different initial mass 
and chemistry. The MC PN populations seem to be ideally suited for these comparisons.

Here we focus on the comparison between AGB stellar evolution and nucleosynthesis models 
and PNe observed in the SMC. Our goals are both the 
identification of particular PN populations to be associated with specific AGB processes,  
and the refinement in our understanding of the main physical aspects of AGB evolution. 
This analysis is complementary to the interpretation of observed samples of AGB stars, 
which is based on the straight comparison between the evolutionary sequences through the 
AGB and the sources observed.

In the first paper of this series (Ventura et al. 2015b, hereinafter paper I), we interpreted the PNe 
observed in the LMC in terms of mass, formation epoch, and chemical composition of their 
progenitors, based on the comparison of observational data with the evolutionary AGB models. Paper I 
confirmed that the models used to reproduce the Spitzer observations of evolved 
stars in the LMC \citep{flavia15a}
could also account for the chemical composition of observed LMC PNe.

The collection of the SMC PNe observational data is presented in Section 2. In Section 3 
we give the main physical and
chemical input adopted to calculate the AGB evolutionary sequences used for the
interpretation of the chemical composition in SMC PNe. The most relevant AGB
evolutionary concepts are discussed in Section 4, while the comparison between
models and observational data is given in Section 5. Section 6 describes the
implications of the present study for some of the most debated points regarding the
AGB evolution. Finally, we summarize our
conclusions in Section 7.

\section{Data selection of SMC PNe}
\label{obs}
Planetary nebulae in the Magellanic Clouds have been observed for many decades, but it is only 
with the use of {\it HST} that we began to resolve them spatially, since they are about half 
an arc second  in diameter. Once observed with the HST, compact HII regions at this 
distance are spatially resolved, and multiple stars that ionize the nebula are evident. 
We selected these compact HII regions out of our PN sample, since contamination of the 
PN sample with HII regions would be misleading for the comparison with AGB stellar 
evolution models that we are endeavoring. It turns out that it is even more likely to 
misclassify PNe in a low metallicity ambient such as the SMC than in other 
environments (Stanghellini et al. 2003a).

The observational sample analyzed in this paper and  presented in Table 1 includes all 
and only the {\it HST} observed and spectroscopically-confirmed SMC PNe whose at least one abundances among 
those of He, C, N, O, Ne is known. In column (1) of Table 1 we list the usual name of 
the PN, where J, MA, MG, SMP, and SP indicate the original discovery or listing catalogs, 
with {\it J} referring to Jacoby (1980), {\it MA} to Meyssonnier \& Azzopardi (1993),  
{\it MG} to  Morgan \& Good (1985), {\it SMP} to  Sanduleak et al. (1978), and  {\it SP} 
to Sanduleak \& Pesch (1981). Note that in some cases they may be listed in two or more 
catalogs, and we use the most commonly used names.

Column (2) lists the PN morphological class. All morphological types are based 
on {\it HST} observations, either using FOC or PC1 (SMP03, SMP05, SMP10, SMP15, SMP16, 
SMP21, Stanghellini et al. 1999) or STIS (Stanghellini et al. 2003) images. Morphologies 
derived from STIS images were preferred when both STIS and one of the pre-refurbishment 
images were available. The morphological codes used in Table 1 have the following meanings: 
{\it R} is for Round, {\it E} for elliptical, {\it EBC} or {\it RBC} for round or elliptical 
with a bipolar core, and {\it B} for bipolar PNe. In some cases, round or elliptical PNe 
have some inner structure {\it (Es)}  or an attached outer halo {\it(ah)}, as indicated 
in the Table. Unresolved PNe are indicated with {\it U}.

Columns (3) through (7) of Table 1 give the nebular abundances, in the usual terms of 
(X/H)=log(X/H)+12.  Column (8) gives the references to the abundances, the first (or only) entry
refers to ground-based abundances (He, N, O, Ne) while the second entry, if present, refers to carbon abundances, which is observed from space-based telescopes. 
 All major recombination lines of carbon that are typically observed 
in PNe fall into the UV regime, thus its total abundance is only quantifiable via space observations, and 
thus harder to come by. It is worth noting that Shaw et al. (2010) calculated helium 
abundances with a method by Porter et al. (2005), but soon after the publication of the 
Shaw et al (2010) paper, another prescriptions by Porter et al. had become available, and an 
erratum was published. To be on the safe side,  where the helium emission lines were 
available from Shaw et al. (2010) we preferred to recalculate them with the 
prescription by Benjamin et al. (1999) to make the data set homogeneous. 

The last two columns of Table 1 give the dust type of the PNe, and the relative reference. These types have been derived from 
{\it Spitzer} IRS spectra in the 4-38 $\mu$m range (Stanghellini et al. 2007, and Bernard-Salas et al. 2008). For the scope 
of this paper it is sufficient to know whether the dust present in the PN has C-rich 
features, O-rich features, or it is featureless (F), as noted in the table. 

Unless otherwise stated, we use the data of Table 1 for all the models-data comparison 
discussion and plots in the present paper.

\begin{table*}
\caption{SMC PN data}                                       
\begin{tabular}{l l c c c c l l l l}        
\hline\hline                        
Name   &   morph.   &   (He/H)  &   (C/H)  &   (N/H)  &   (O/H)  &   (Ne/H)  &   Ab. ref.$^a$&  Dust  type& dust ref.$^b$    \\
(1)&(2)&(3)&(4)&(5)&(6)&(7)&(8)&(9)&(10)\\
\hline       
     J~04 &      E & 11.02 & $\dots$   &  8.04 &  7.30 &  6.76 &   SRM&    $\dots$&  $\dots$ \\
      J~18 &      R$^d$ & 10.30 & $\dots$   &  8.04 &  7.17 &  6.67 &   SRM&     $\dots$ &  $\dots$\\
      J~23 &      U & 11.03 & $\dots$   & $\dots$ &  6.33 &  6.96 &   SRM&     $\dots$ &  $\dots$\\
      J~27 &      B$^d$ & $\dots$ & $\dots$   &  7.97 &  7.86 & $\dots$ &   KJ&    $\dots$ &  $\dots$\\
   MA~1682 &      B & 11.19 & $\dots$   & $\dots$ &  7.40 &  7.34 &    SRM&    $\dots$ &  $\dots$\\
   MA~1762 &    EBC & 10.93 & $\dots$   & $\dots$ &  7.29 &  6.39 &    SRM&    $\dots$&  $\dots$ \\
     MG~08 &     Es & 11.08 & $\dots$   &  7.15 &  8.15 &  6.39 &   S10$^c$&    $\dots$&  $\dots$ \\
     MG~13 &     Es & 10.94 & $\dots$   &  5.12 &  8.00 &  7.12 &    S10$^c$&   $\dots$ &  $\dots$\\
     SMP~1 &      U & 10.83 & $\dots$   &  7.16 &  7.86 &  6.42 & LD&  C-rich & BS09\\
     SMP~2 &      R & 11.11 &  8.74   &  7.47 &  8.01 &  7.21 & LD, LDL& C-rich &S07\\
     SMP~3 &      B & 10.94 &  8.77   &  7.00 &  7.98 &  7.00 &  LD, LD& C-rich  & BS09\\
     SMP~5 &      R & 11.11 &  8.90   & $\dots$ &  8.24 & $\dots$ &  LD, LDL& C-rich &S07\\
     SMP~6 &      E & 10.95 &  8.35   &  7.71 &  8.24 &  7.38 &  LD96, S09& C-rich  & BS09\\
     SMP~8 &      R & 11.03 &  8.12   &  7.05 &  7.88 &  7.37 &   S10$^c$, S09&     F &S07\\
     SMP~9 &     Rs & 10.85 &  8.79   &  7.25 &  8.32 &  7.51 &   S10$^c$, LD&     F &S07\\
    SMP~10 &      B & 10.85 &  9.82$^e$   &  7.82 &  8.58 &  7.65 & LD, LD96&      $\dots$ &  $\dots$\\
    SMP~11 &      B & 11.16 & $\dots$   &  6.52 &  8.02 &  6.90 & S10$^c$&  C-rich  & BS09\\
    SMP~12 &      E & 11.01 & $\dots$   & $\dots$ & $\dots$ &  6.89 & LD&      $\dots$ &  $\dots$\\
    SMP~13 &      R & 11.05 &  8.73   &  7.30 &  8.06 &  7.35 &  S10$^c$, S09& C-rich &S07\\
    SMP~14 &      R & 10.94 & $\dots$   &  7.36 &  8.29 &  7.65 &  S10$^c$, LD& C-rich&S07 \\
    SMP~15 &      R & 11.14 &  8.26   &  7.38 &  8.21 &  7.67 &  LDL, S09& C-rich &S07\\
    SMP~16 &      E & 10.69 &  8.19   &  6.55 &  7.85 &  6.37 &  LD, S09& C-rich&S07 \\
    SMP~17 &  R(ah) & 11.05 &  8.40   &  7.50 &  8.24 &  7.58 &  S10$^c$, LDL& C-rich&S07 \\
    SMP~18 &      U & 11.04 &  8.31   &  7.11 &  7.90 &  7.57 &  S10$^c$, S09&C-rich&S07 \\
    SMP~19 &      R & 11.00 &  8.97   &  7.28 &  8.14 &  7.63 &  S10$^c$, LD& C-rich&S07 \\
    SMP~20 &      U & 11.07 &  8.25   &  6.95 &  7.74 &  6.91 &  S10$^c$, S09& C-rich&S07 \\
    SMP~21 &     Es & 11.00 &  7.13   &  7.92 &  7.55 &  6.84 &   LD, LD96&   $\dots$ &  $\dots$\\
    SMP~22 &      B$^d$ & 11.07 &  7.23   &  8.05 &  7.59 &  6.72 &    LD96, LD&   F& BS09 \\
    SMP~23 &    EBC & 11.00 &  8.39   &  7.24 &  7.93 &  7.37 &    S10$^c$, LD&   F &S07\\
    SMP~24 &      E & 11.04 &  8.18   &  7.17 &  8.06 &  7.36 &  S10$^c$, S09& C-rich&BS09 \\
    SMP~25 &      E & 11.02 &  6.64   &  7.92 &  7.56 &  6.96 &  LD, S09& O-rich &S07\\
    SMP~26 &      P & 10.96 &  8.46   &  8.10 &  8.14 &  7.43 &   LD, S09&     F &S07\\
    SMP~27 &  R(ah) & 11.04 & $\dots$   &  7.17 &  8.00 &  7.20 &  S10$^c$& C-rich&S07 \\
    SMP~28 &      E & 11.11 &  6.96   &  8.04 &  7.46 &  6.85 &   LD96, S09&     F&BS09 \\
     SP~34 &  R(ah) & 10.86 & $\dots$   &  7.30 &  7.82 &  7.05 &    SRM&   $\dots$&  $\dots$ \\

 \hline     
               
\end{tabular}

Table reference notes. 
$^a$ Abundance reference keys: KJ (Kaler \& Jacoby 1980); 
LD (Leisy \& Dennefeld 2006); LD96 (Leisy \& Dennefeld 1996); LDL are from LD96 with 
lines from the literature cited therein; S10  (Shaw et al. 2010); SRM 
(Stasinska et al. 1988); S09 (Stanghellini et al. 2009).  
$^b$ Dust references keys:
BS09 (Bernard-Salas et al. 2009); S07  (Stanghellini et al. 2007).  $^c$(He/H) recalculated 
from S10 emission lines and Benjamin et al. (1999) prescription (see text).
$^d$ Uncertain morphology. 
$^e$ Uncertain abundance.            

\end{table*}

\section{Modelling AGB evolution and dust formation}
\subsection{Stellar evolution models}
The evolutionary sequences used in this work were calculated with the ATON code; an interested
reader can find the details of the numerical and physical structure of the code in
\citet{ventura98}, whereas the most recent updates are presented in \citet{ventura09}.

The ingredients used most relevant for the present investigation are the following:

\begin{itemize}
\item{
{\it Chemical composition}. The AGB models are calculated in the mass range
$0.8~\Msun \leq M \leq 8~\Msun$\footnote{The upper limit to the AGB evolution is indeed
slightly dependent on the metallicity. For low metallicity models, i.e. $Z=1,2\times 10^{-3}$,
the highest mass considered is $7.5~\Msun$. Models with mass $M\geq 7~\Msun$ develop 
a core composed of oxygen and neon.} and for the metallicities $Z=1,2,4,8 \times 10^{-3}$.  
The mixture adopted is taken from \citet{gs98}. In agreement 
with our previous investigations, for the $Z=1,2 \times 10^{-3}$
models we used an $\alpha$-enhancement $[\alpha/Fe]=+0.4$, whereas for the higher Z
models we used $[\alpha/Fe]=+0.2$; in the interpretation of the results, we will take into
account that the best choice for the SMC stars would be
a solar-scaled or even sub-solar distribution of $\alpha$ elements. A few test models
with $[\alpha/Fe]=-0.2$ for the $Z=2 \times 10^{-3}$ metallicity were calculated.
The initial chemical composition for the four sets of models used here is reported in 
Table 2, in terms of the CNO elements and of helium.
}

\item{
{\it Convection.} The temperature gradient within  
regions unstable to convective motions is found by means of the full spectrum of turbulence
(FST) model \citep{cm91}. Nuclear burning and mixing of chemicals
are coupled in a diffusive-like scheme \citep{cloutmann}. 
During the AGB phase, overshoot of convective eddies from the bottom of the envelope and
from the top and bottom borders of the shell which forms at the ignition of each 
thermal pulse,
is described by means of an exponential decay of velocities from the convective/radiative
interface, fixed by the Schwarzschild criterion. The e-folding distance is assumed to
be $0.002H_p$ (where $H_p$ is the pressure scale height calculated at the formal boundary
of convection), in agreement with the calibration based on the observed luminosity
function of carbon stars in the LMC, given by \citet{paperIV}. 
}

\item{
{\it Mass loss.} The mass loss rate for oxygen-rich models is determined via the 
\citet{blocker95} treatment; following the calibration based on the luminosity
function of carbon stars in the LMC by Ventura et al. (2000), we set the free parameter 
entering the \citet{blocker95}'s recipe to $\eta_R=0.02$. For $C/O>1$ environments we used 
the results from the Berlin group \citep{wachter02, wachter08}.
} 

\item{{\it Molecular opacities.} The molecular opacities in the low-temperature regime (below 
$10^4$ K) are calculated by means of the AESOPUS tool \citep{marigo09}. 
The opacities are suitably constructed to follow the changes in the chemical composition 
of the envelope, particularly of the individual abundances of carbon, nitrogen, and oxygen.
} 

\end{itemize}

\begin{table}
\caption{Initial chemical composition of the AGB models}                                       
\begin{tabular}{c c c c c c}        
\hline\hline                        
Z  &   (He/H)  &   (C/H)  &   (N/H)  &   (O/H)  &   (Ne/H)   \\
\hline       
$10^{-3}$ & 11.52 & 8.05 & 7.52 & 8.89 & 8.23         \\
$2\times 10^{-3}$ & 11.52 & 8.35 & 7.82 & 9.19 & 8.53 \\
$4\times 10^{-3}$ & 11.55 & 8.82 & 8.29 & 9.46 & 8.80 \\
$8\times 10^{-3}$ & 11.55 & 9.13 & 8.60 & 9.76 & 9.11 \\
 \hline     
               
\end{tabular}
\end{table}

\subsection{Dust formation in the winds of AGB stars}
The formation and growth of dust particles in the wind of AGB stars is described by
means of the pioneering scheme described by the Heidelberg group \citep{fg06}. All
the relevant equations, with a detailed explanation of the physical assumptions
adopted, can be found in \citet{fg06} and in the series of papers on this argument
published by our group \citep{paperI, paperII, paperIII, paperIV}.

The wind is assumed to expand isotropically from the surface of the star. Mass and
momentum conservation allows the determination of the radial profiles of density and
velocity. The effects of dust on the dynamics of the wind is described by the
extinction coefficient, $k$, which is used to calculate the radiation pressure and the
consequent acceleration of the wind; this quantity depends on the number density and size
of dust grains formed (eqs. 2 and 9 in Ferrarotti \& Gail 2006).

The dust species formed depend on the chemical composition of the surface layers of the
stars, primarily on the C/O ratio. For oxygen-rich environments we consider the formation 
of silicates and alumina dust, whereas in the wind of carbon stars we account for the 
formation of solid carbon and silicon carbide. The rate of the grain growth is found on 
the basis of the values of density and of the mass fraction of the chemical elements 
relevant for the formation of the various dust species, namely silicon (silicates and 
silicon carbide), aluminium (alumina dust) and carbon (solid carbon).

The formation of dust has a direct influence on the AGB evolution, because it favours
the acceleration of the wind, thus an increase of the rate with which mass is lost from
the envelope. 

\section{The AGB evolution}

The models presented here are extensively illustrated and discussed in \citet{ventura14b} 
($Z=4\times 10^{-3}$), \citet{ventura13} ($Z=1,8\times 10^{-3}$, initial mass above 
$3~M_{\odot}$) and \citet{paperIV} (low--mass models of metallicity $Z=1,8\times 10^{-3}$ 
and initial mass below $3~M_{\odot}$). The models are the same used in paper I, with the
only exception of the $Z=2\times 10^{-3}$ evolutionary sequences, which were calculated
specifically for the present investigation.

The evolution of the surface chemical composition of AGB stars is known to be driven by the combined effects
of TDU and HBB \citep{karakas14}. The former favours an increase in the surface carbon and, in smaller quantities,
of the oxygen content. HBB provokes the change in the relative fractions of the
CNO elements, according to the equilibrium determined by the temperature at the base of
the convective envelope. The surface carbon is destroyed in favour of
nitrogen; for temperatures above $\sim 80$MK oxygen depletion occurs.
TDU is the dominant mechanism for stars of initial mass below $\sim 3~\Msun$, whereas
the effects of HBB prevail in the high-mass regime \citep[see, e.g.][]{ventura13}.

\begin{figure*}
\begin{minipage}{0.33\textwidth}
\resizebox{1.\hsize}{!}{\includegraphics{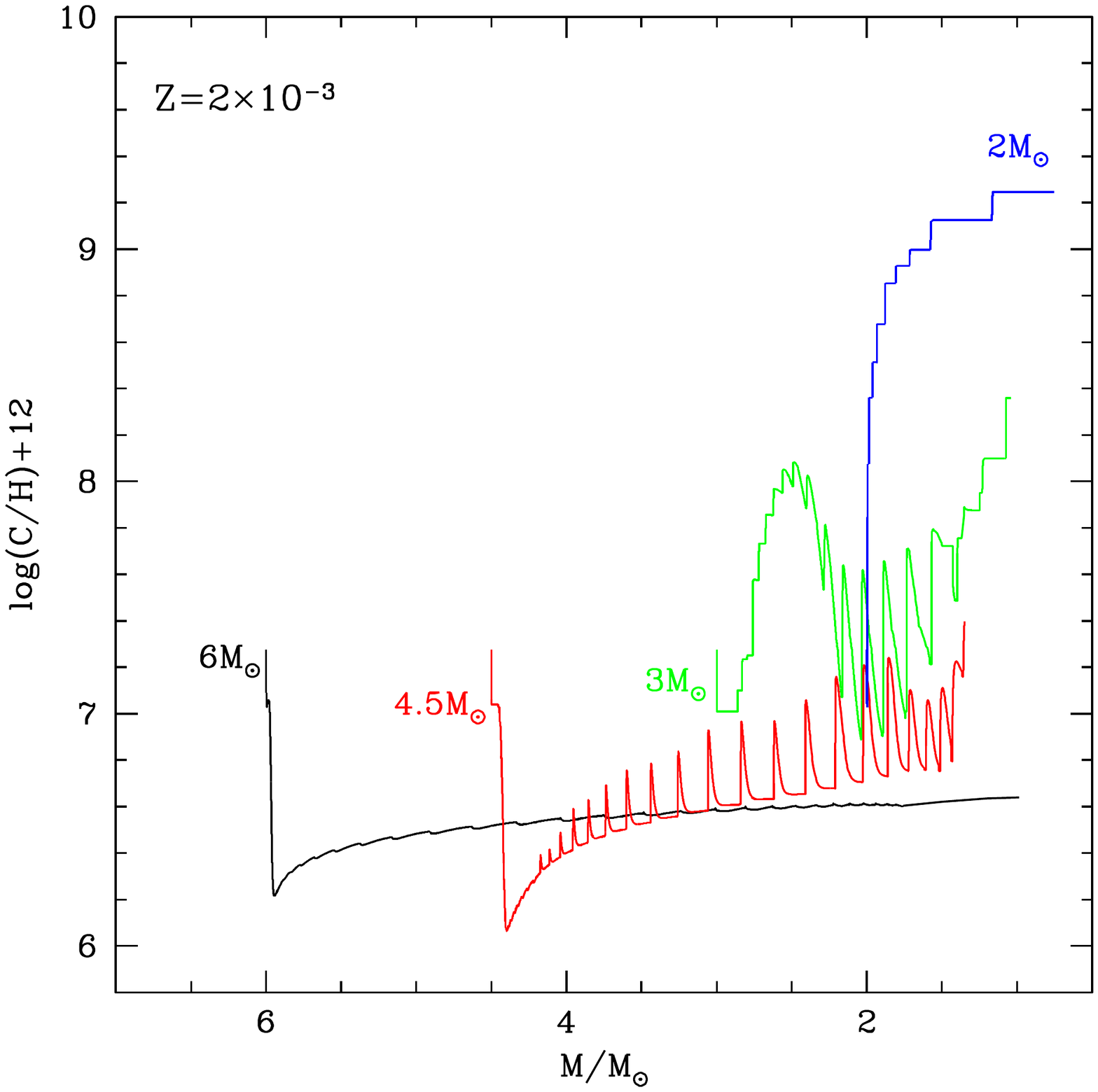}}
\end{minipage}
\begin{minipage}{0.33\textwidth}
\resizebox{1.\hsize}{!}{\includegraphics{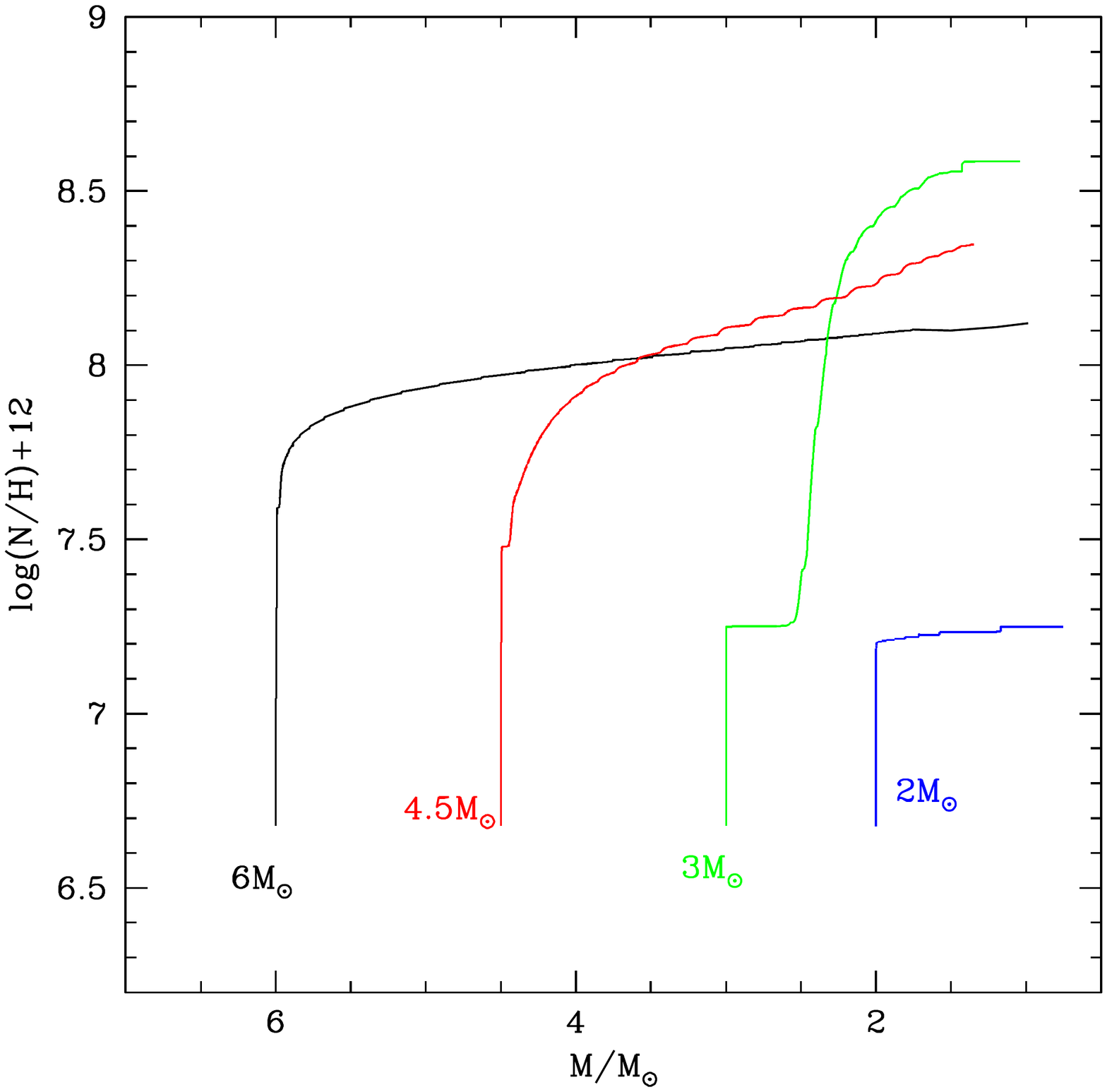}}
\end{minipage}
\begin{minipage}{0.33\textwidth}
\resizebox{1.\hsize}{!}{\includegraphics{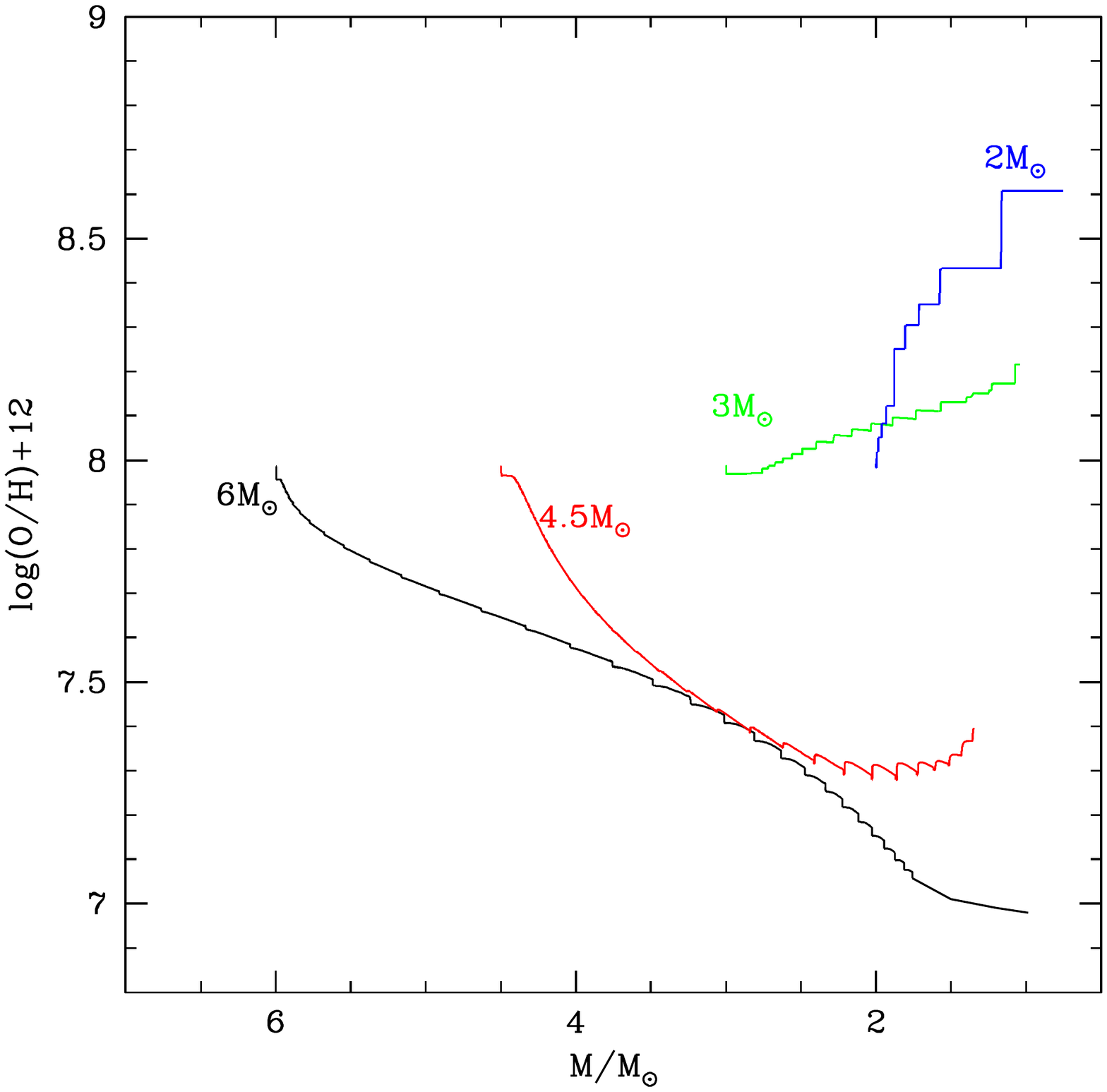}}
\end{minipage}
\vskip-30pt
\caption{The evolution of the surface mass fraction of carbon (left), nitrogen 
(middle) and oxygen (right) during the AGB phase of models of metallicity 
$Z=2\times 10^{-3}$ and of initial masses $2$, $3$, $4.5$ and
$6~M_{\odot}$.
}
\label{fcno}
\end{figure*}

Fig.~\ref{fcno} shows the evolution of $Z=2\times 10^{-3}$ models of different  mass
in terms of the individual mass fractions of carbon, nitrogen and oxygen. Similar plots
for the other metallicities were shown in paper I. To allow the simultaneous plot of all 
the evolutionary sequences, we use the stellar mass as abscissa instead of the AGB time;
the individual sequences must be read rightwards, as mass diminishes during the evolution. 
The models shown in the figure can be considered as representative of the different 
behaviours.

\begin{enumerate}
\item{
Stars of mass below $\sim 3~\Msun$ (represented by the $2~\Msun$ model in
Fig.~\ref{fcno}) do not experience any HBB; consequently, the
variation of the chemical composition is entirely determined by TDU. For stars of initial
mass above $\sim 0.9~\Msun$ the final chemistry is enhanced in carbon and, in more 
modest quantities, in oxygen: these stars reach the carbon star stage during the AGB 
evolution and end up their life with a surface C/O above unity. The nitrogen content is 
changed after the first dredge-up episode during
the red giant branch phase and remains practically unchanged for the whole AGB phase.
Among the stars belonging to this group, the model shown in Fig.~\ref{fcno} is one of
those undergoing the greatest variation of the surface chemical composition, as result
of the effect of several TDU events; in models of smaller mass, the increase in the
carbon and oxygen content is smaller.}

\item{The chemical composition of stars of mass $3~\Msun < M < 4~\Msun$ 
is affected by both TDU and HBB.  
These objects are the most efficient manufactures of nitrogen, because the latter species 
is produced not only by the carbon initially present in the envelope, but also by the 
additional carbon dredge-up after each thermal pulse. In Fig.~\ref{fcno} this class of
objects is represented by the $3~\Msun$ model. The variation of the surface carbon is
the most indicative of the balance between TDU and HBB: the increase of C in the initial
phases are a consequence of TDU, whereas the following depletion of carbon is due to the
HBB activity. In the very final AGB phases, when HBB is extinguished, the C-star stage
can be achieved. 
}

\item{HBB is the dominant mechanism in changing the surface chemistry of stars with
mass in the range $4~\Msun < M < 6~\Msun$. However, these objects are also expected to
experience a few TDU episodes towards the end of the AGB evolution, when HBB is
extinguished. These stars are efficient nitrogen producers, whereas their carbon and 
oxygen content is smaller compared to the initial abundance. The final carbon content
depends on the number of TDU episodes experienced during the latest thermal pulses.
In Fig.~\ref{fcno} we represent this group of stars by showing the evolution of
a $4.5~\Msun$ model. The effects of HBB can be seen in the initial decrease in the surface
carbon and in the drop in the carbon mass fraction which follow each thermal pulse,
after the increase due to TDU.
}

\item{The chemical composition of massive AGB stars, with mass $M > 6~\Msun$, reflects
the pure effect of HBB. The overall effect of the nuclear activity at the bottom of the
convective envelope is the reduction of the surface carbon and oxygen, and the increase in
the nitrogen mass fraction; the overall C+N+O content is unchanged. The $6~\Msun$ model
in Fig.~\ref{fcno} was chosen as a representative of this class of objects. The effects
of HBB can be seen in the initial drop in the surface carbon and in the decrease in
the surface oxygen, which occurs in the following evolutionary phases. Note that the
surface oxygen is much more sensitive than carbon to the temperature at which HBB occurs:
looking at the $6~\Msun$ tracks in Fig.~\ref{fcno}, we see that while the oxygen decrease
continues during the whole AGB phase, as a consequence of the higher HBB temperatures,
the surface carbon, after the initial decrease, keeps approximately constant. 
}
\end{enumerate}

\begin{figure*}
\begin{minipage}{0.49\textwidth}
\resizebox{1.\hsize}{!}{\includegraphics{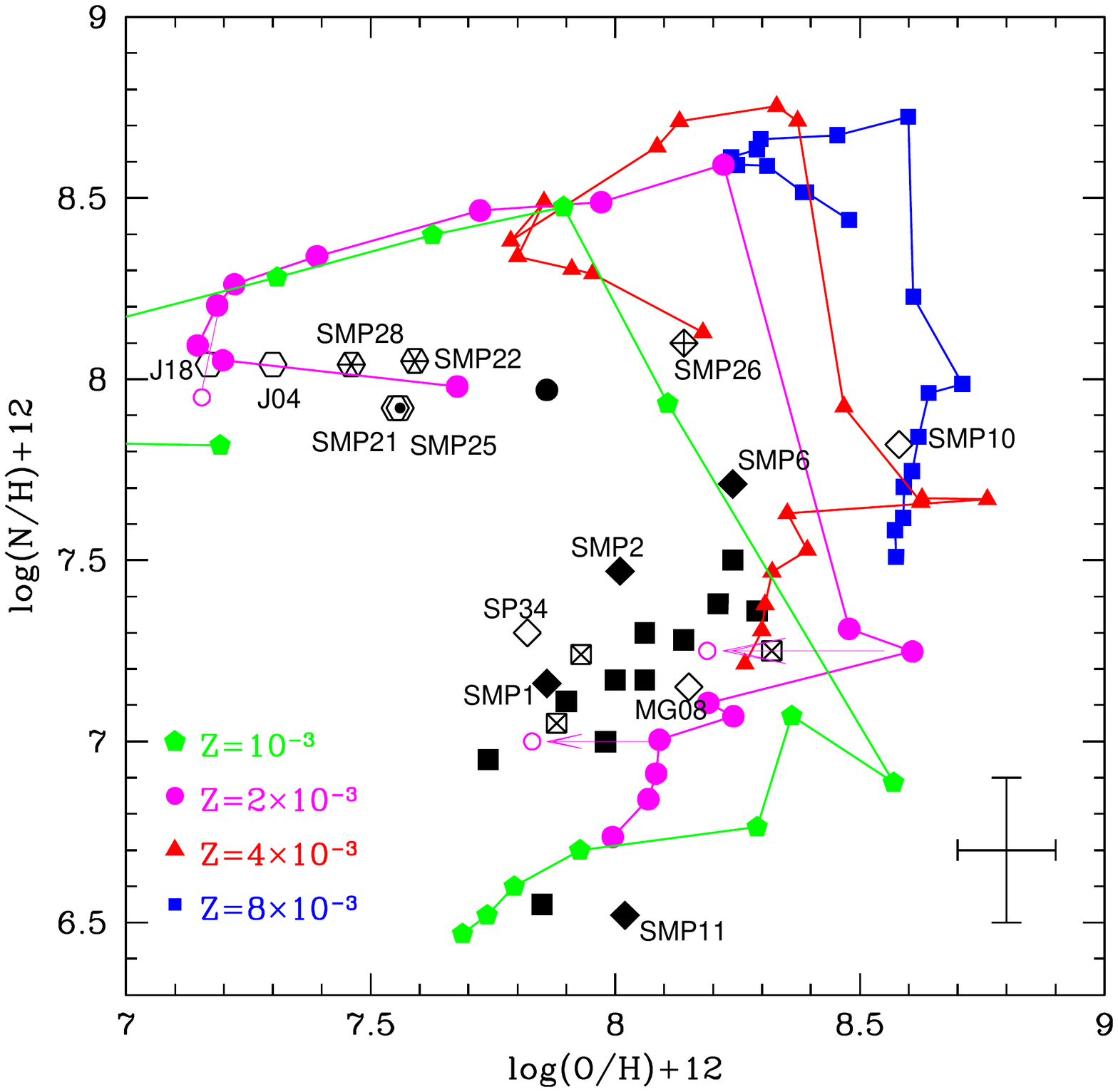}}
\end{minipage}
\begin{minipage}{0.49\textwidth}
\resizebox{1.\hsize}{!}{\includegraphics{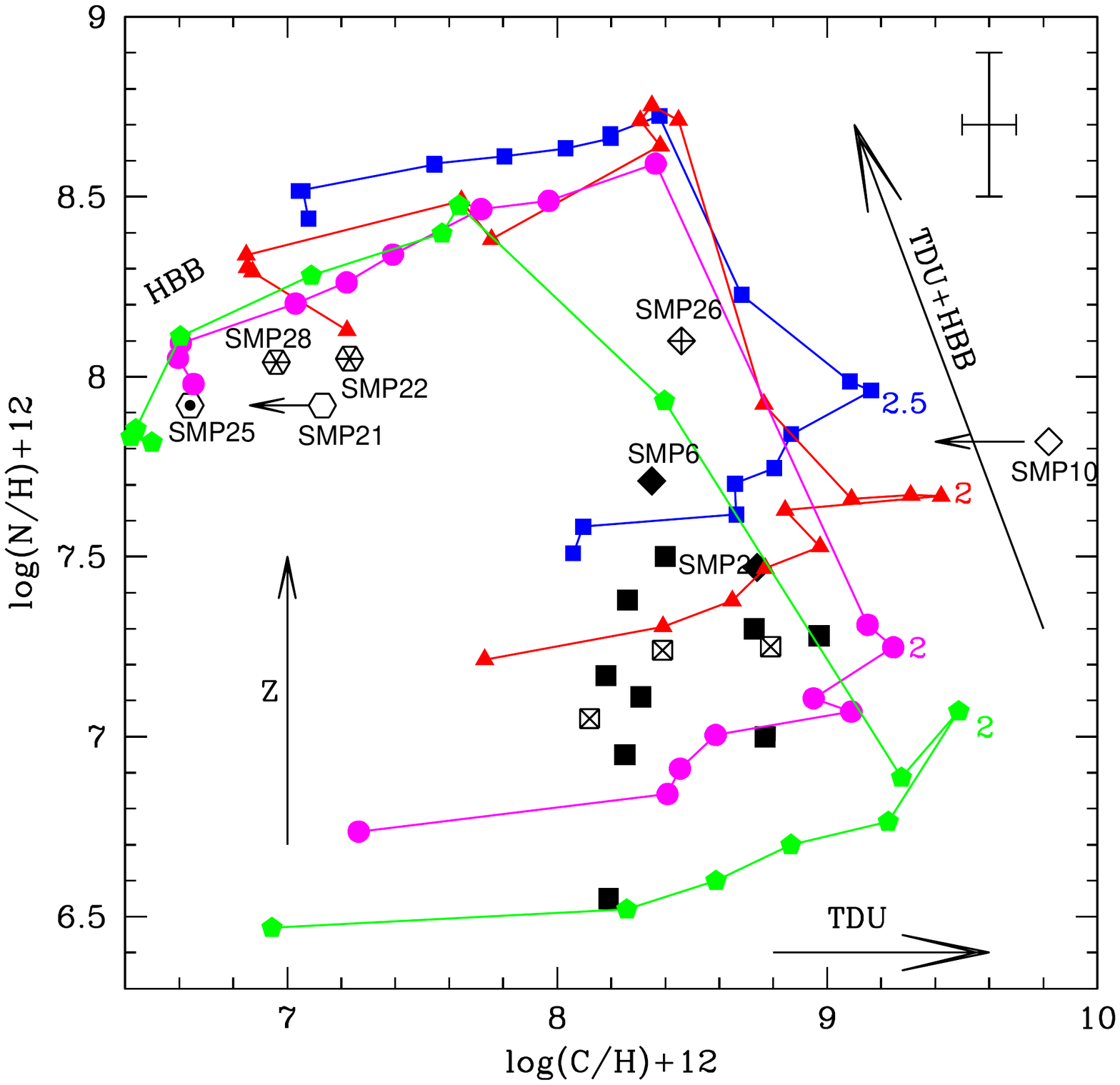}}
\end{minipage}
\vskip-50pt
\caption{The observed chemical composition of the SMC PNe sample in
the O-N (left panel) and C-N (right) planes. Squares, hexagons, and diamonds indicate the 
loci of PNe respectively of groups (a), (b) and (c)
(see $\S$5). Filled symbols mark PNe with traces of carbon dust, open-dotted
symbols indicate evidence of silicates, open symbols indicate PNe 
with no indication of dust in the literature, and crossed symbols indicate featureless IRS
spectra. We indicate the PN names for groups (b) and (c) PNe. 
Typical errors associated to the determination of the abundances of the individual species
are indicated with crosses in the right, bottom side of the left panel and in the
right, top region of the right panel.
The final abundances of the AGB models of metallicity $Z=10^{-3}$ (green pentagons), 
$Z=2\times 10^{-3}$ (magenta circles), $Z=4\times 10^{-3}$ (red triangles), 
$Z=8\times 10^{-3}$ (blue squares) are also shown (see legend in the O-N plot). 
Open, magenta circles in the left panel refer to $Z=2\times 10^{-3}$
models calculated with a lower initial oxygen ($[O/Fe]=-0.2$).
}
\label{fpne}
\end{figure*}

The range of masses of the four groups given above are slightly dependent on the
metallicity of the models. This holds in particular for the first and the second group, 
because HBB is activated more easily in stars of lower metallicity: therefore, the upper 
limit of the mass of the stars experiencing solely TDU changes from $\sim 3~\Msun$ for 
$Z=4,8\times 10^{-3}$, down to $\sim 2.5~\Msun$ for $Z=1,2\times 10^{-3}$. Concerning
group (iv), the highest mass not undergoing core collapse is
$\sim 8~\Msun$ for $Z=4,8\times 10^{-3}$, whereas it is $\sim 7.5~\Msun$ for 
$Z=1,2\times 10^{-3}$.

The final chemical composition of the various models used in the present work are shown
in Fig.~\ref{fpne}, in the C vs N and O vs N planes. Tracks of different colors 
connect models of different metallicity and different masses in the range 
$0.9~\Msun < M < 8~\Msun$, where masses increase counter-clock wise for each metallicity 
track. In the low-mass regime, corresponding to the stars discussed in point (i) above, 
the final carbon and oxygen increase
with the initial mass of the star, because more massive objects undergo a higher number
of TDU events. The trend in both planes is reversed for masses around $3~\Msun$
(point ii above), because HBB sets in, determining the increase in the surface nitrogen
and a partial destruction of the carbon accumulated by TDU. In the higher mass domain
(points iii and iv above) the theoretical sequences gradually move to the left-upper regions
of the diagrams, those corresponding to a chemical composition entirely determined by
the equilibria of HBB.

The metallicity of the star affects the final chemical composition owing to the
difference in the initial mass fractions of the relevant dust species, and to the
sensitivity of some of the mechanisms able to alter the surface chemical composition
to the metal content of the star. We stress here two main effects of metallicity:

\begin{itemize}

\item{In the low-mass domain, the chemical species mostly affected by changes in the 
metallicity is nitrogen. The final N of these objects is solely determined by the 
first dredge-up event, taking place during the red giant branch evolution. Therefore, 
a spread in the metallicity of the stars will reflect into a spread in the surface N.
}

\item{Concerning massive AGBs, the stars of smaller metallicity undergo stronger HBB, 
with higher temperatures at the base of the convective envelope \citep{ventura13};
therefore, low-Z models experience a more advanced nucleosynthesis. This will reflect 
essentially in a smaller oxygen, because the higher temperatures lead to a more efficient 
activation of the full CNO cycling; conversely, in higher-Z model little depletion of 
oxygen occurs. In agreement with the arguments of point (iv) before, we find that
the final oxygen is the best metallicity indicator in these stars. This
is because carbon burning requires smaller temperatures, thus it is activated in models
of all metallicities; furthermore, the carbon equilibrium abundance is not very sensitive
to the temperatures at which HBB occurs, once the minimum threshold required to ignite
HBB, i.e. $T \sim 40$MK, is reached.
}

\end{itemize}

\section{Planetary nebulae in the SMC}

In order to interpret each PN in our sample to the evolutionary path that lead to the 
observed configuration we use the comparison between the nebular abundances and the AGB 
yields of the final ejected envelopes. In the panels of Fig.~2 we show the data points in 
black, and the model points in colors. The final abundances of the AGB models of different 
initial metallicity, given in Table 2, are given by different symbols and color (see the 
legend on the left panel). The left panel, showing N vs. O abundances, and the right panel, 
showing N vs. C abundances, show that models encompass well the observed PNe. Before we 
describe in detail the different PN groups of Fig. 2 let us examine closely the metallicity 
dependence of the various populations.

\begin{figure}
\resizebox{1.\hsize}{!}{\includegraphics{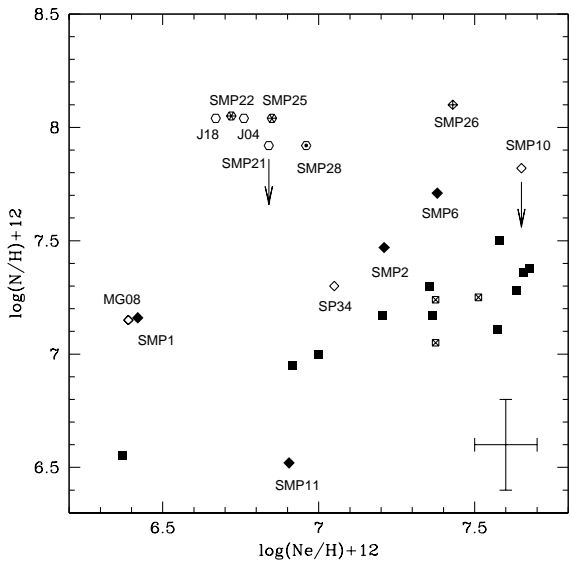}}
\vskip-60pt
\caption{The nitrogen abundances of the PNe sample introduced in section \ref{obs}
as a function of their neon content. The symbols are the same as in Fig.~\ref{fpne}.
The cross in the right, lower part of the plane gives the typical errors in the
neon and nitrogen abundances.
}
\label{fnen}
\end{figure}

\begin{figure}
\resizebox{1.\hsize}{!}{\includegraphics{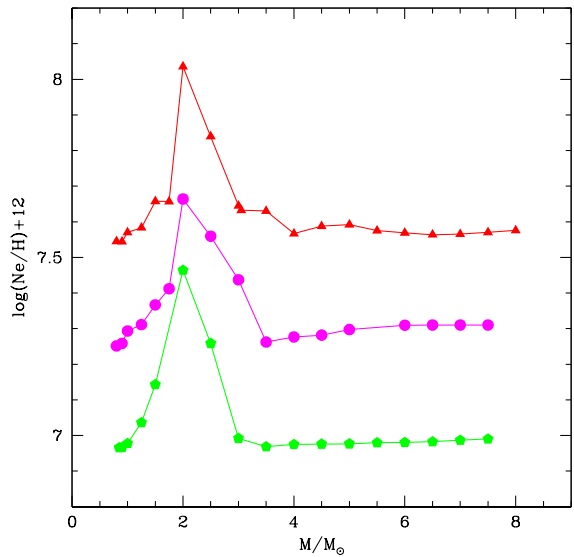}}
\vskip-60pt
\caption{The final neon abundance of the models used in the present analysis
as a function of the initial mass. The metallicities $Z=10^{-3}$, $Z=2\times 10^{-3}$
and $Z=4\times 10^{-3}$ are indicated, respectively, with green pentagons,
magenta circles and red triangles. The initial abundances of neon for the three cases
are $\log(Ne/H)+12=6.93$ ($Z=10^{-3}$), $\log(Ne/H)+12=7.24$ ($Z=2\times 10^{-3}$), 
$\log(Ne/H)+12=7.51$ ($Z=4\times 10^{-3}$)
}
\label{fneon}
\end{figure}

In Fig. 2, left panel, there is a notable sequence of PNe where N and O are almost 
correlated. To further investigate the relationship between nitrogen and the metallicity,
we show the N against Ne trend in Fig.~\ref{fnen}. We prefer neon to argon and sulphur as 
metallicity indicator of the PNe progenitors, because the neon abundance is available for 
all but two of the PNe considered here and, more important, the errors associated to the 
measurements of neon are significantly smaller in comparison to argon and 
sulphur \citep{leisy06}. An additional reason to disregard sulphur is the well known
"sulphur anomaly" (Henry et al. 2012); e.g., it may even be depleted into dust 
\citep{pottasch06}. The only problem with using neon as metallicity indicator is
that the surface abundance of this chemical species can be changed during the AGB phase:
however, as shown in Fig.~\ref{fneon}, the only significant changes, at most by a factor
$\sim 3$, are obtained in stars in a narrow range of mass around $\sim 2\Msun$ (i.e. those 
undergoing the most relevant modification via TDU). To further test the relationship
between nitrogen and metallicity, we checked the argon - nitrogen trend and verified
that it follows the same behaviour, though with a higher dispersion, as the Ne-N relationship.

The majority of PNe in Fig.~\ref{fnen} follow an 
approximately linear Ne-N trend, indicating that N increases with metallicity; we will 
refer to this PN sequence as group (a) in the following, and plot them with square symbols 
in the figures. Notably displaced from the sequence, with large enhanced nitrogen, there 
is a smaller group of PNe which we plot with hexagons in the figures and refer as group (b). 
Finally, we can identify another minor group of PNe, indicated with diamonds in the figures, 
whose position in the plane deviates slightly (more than 0.3 dex) from the main N-Ne 
trend traced by group (a) PNe. We will refer to this subsample as group (c). All the PNe  
in group (a) and (c) exhibit C/O ratios above unity.

\subsection{Carbon-rich PNe}
In the right panel of Fig.~\ref{fpne} we see that all group (a) PNe, indicated with 
squares, have enhanced carbon. We interpret these PNe as the progeny of stars of mass 
$M < 2~\Msun$. The observed spread in N of group (a) PNe is due to differences in the 
progenitor's  metallicity of these objects, as clearly shown in Fig.~\ref{fnen}. Most of 
these PNe descend from $\sim 1-1.5~\Msun$ stars of metallicity in the range
$2\times 10^{-3} < Z < 4\times 10^{-3}$, formed between 1.5 Gyr and 7 Gyr ago. 
PNe with  $\log (C/H)+12 > 8.5$ can be 
interpreted as the final phases of stars of higher mass ($\sim 2-2.5~\Msun$), formed in 
more recent epochs, $0.5-1$ Gyr ago. The models nicely reproduce the
observed spread in carbon; the only exception is PN
SMP10, where only upper limit carbon abundance is available in the literature \citep{leisy06}.
The Spitzer IR spectra of most group (a) SMC PNe
exhibit C-rich dust features, indicated with filled symbols in the plots. This is consistent 
with the evolution through the C-rich phase, already noted by Stanghellini et al. (2007). 
Conversely, no PN with traces of O-rich dust in their Spitzer spectra is found in group (a).

Oxygen abundances of group (a) PNe show a considerable offset to lower abundances with 
respect to the Z=$2-4\times$10$^{-3}$ models (left panel of Fig.~\ref{fpne}),
indicating that the original composition of SMC interstellar material is oxygen-poor with 
respect to what has been assumed for the models (see Table 2). This is confirmed by a few 
test $Z=2\times 10^{-3}$ models, calculated with a lower-O mixture, that nicely reproduce 
several of the observed points of group (a). This result is not surprising, given the low 
oxygen abundances detected in F supergiant stars in the SMC \citep{spite89} and the chemical 
and structural evolutionary models for the SMC, which predict lower oxygen for stars in 
the SMC, compared to objects of the same metallicity in the LMC and in the Milky 
Way \citep{russell92}.

\subsection{PNe enriched in nitrogen: the signature of HBB}
PNe in group (b) are indicated with Hexagons in Figs. 2 and \ref{fnen}. Their surface chemical composition shows the 
signature of HBB: this is clearly visible in the enhancement of nitrogen and in the
low abundances of carbon and oxygen. The analysis of their loci on the CN plane (Fig.~\ref{fpne}, 
right panel) supports the idea that these PNe descend from massive AGB stars, whose 
metallicity is in the range $10^{-3} < Z < 4\times 10^{-3}$. The ON plane, for the reasons 
given at the end of the previous section, allows a better discrimination among the various 
metallicities, suggesting that only low-Z models are compatible with the observed abundances 
of oxygen. We conclude that PNe in group (b) are the progeny of low metallicity stars of 
mass $6~\Msun < M < 7.5~\Msun$,
formed $40-70$ Myr ago. The chemical composition of these stars is not affected by TDU; 
this confirms the possibility, outlined in paper I, that the chemistry of massive AGB stars
shows the sole chemical imprinting of HBB, with no contamination from TDU. 
Note that this conclusions still holds if we consider models with lower initial oxygen abundance. 
This is shown in the left panel of Fig.~\ref{fpne}, where we compare the PNe with a $6~\Msun$ 
model of metallicity $Z=2\times 10^{-3}$, 
calculated with an $[O/Fe]=-0.2$ mixture. Compared to the $6~\Msun$, $[O/Fe]=+0.2$ model, 
the main difference is that the final N is a factor of $\sim 2$ lower (note that this is in 
better agreement with the data of group (b) PNe), whereas the final O is only slightly
($\sim 0.1$ dex) less abundant. This is because the $[O/Fe]=-0.2$ model has a smaller, 
initial C+N+O and, under HBB conditions, N is the main product of the nucleosynthesis
activated.

Apart from SMP~21, whose carbon abundance is an upper limit and very unreliable, 
there are three clear members of group (b): SMP~22, SMP~25, and SMP~28. Interestingly, 
SMP~25 is the only PN in the SMC that has been detected to have O-rich dust, and this is 
perfectly compatible with having a higher-mass progenitor, as derived from the comparison 
with the tracks. The other two PNe have also {\it Spitzer} spectra available, with {\it F} 
dust spectra, which typically means that their dust has sputtered by the time of 
observations (Stanghellini et al. 2007). This is also consistent with a higher mass 
progenitor and a generally rapid shell evolution.

While on a pure chemical point of view the possibility that PNe in group (b) descend from
massive AGB progenitors is in nice agreement with the observations, statistical arguments
are at odds with the hypothesis that such a large ($\sim 18\%$) fraction of the PNe in 
the sample are the progeny of high mass AGB stars. The short post-AGB lifetimes of stars
with core mass above $0.9\Msun$ ($\sim 60$ yr, Bloecker 1995) and the relatively small
percentage of stars with mass in the range $6-8\Msun$ expected on the basis of any realistic
mass function are marginally compatible with the possibility that 6 out of 35 PNe descend 
from massive AGB progenitors.

Part of group (b) PNe might descend from low-mass stars of initial mass below $\sim 1.25 \Msun$
(thus not becoming C-stars), which experienced some non-standard mixing during the
previous giant branch phases, thus enriching the surface layers in N-rich and C-poor 
material. This hypothesis is supported by results showing that stars of mass below 
$\sim 2~\Msun$ experience deep mixing during the ascent of the first giant branch 
(Gilroy 1989, 1991; Angelou et al. 2012). For half of the stars in this group the low 
oxygen abundances measured are compatible with low metallicity progenitors, if a
0.2 dex error bar is assumed.

A further possibility is that some PNe in group (b) belong to binary systems
composed of low-mass stars, which evolved through a common envelope phase at the tip
of the RGB or during the early AGB phase; these systems would show the imprinting
of CN cycling, avoiding the achievement of the C-star phase. On the statistical side,
this evolutionary scenario is plausible, given that $\sim 50\%$ of stars with mass above 
$1\Msun$ are part of binary systems.

\subsection{A few outliers}
Group (c) PNe, plotted with diamonds in the Figures, exhibit anomalous
nitrogen abundances, falling off the main N-Ne locus in Fig.~\ref{fnen}. 
Planetary nebulae SMP~1, SMP~2, SMP~11, SP~34, and MG~8 populate the region in the ON 
plane reproduced by low-mass stars models (see left panel of Fig.~\ref{fpne}), 
which became carbon stars after experiencing a few thermal pulses. We are forced to base 
our analysis on this plane only, because the carbon abundance is available only for the 
source SMP~2. The N abundance is so small that any HBB effect can be disregarded.
We conclude that their anomalous nitrogen is likely
due to intrinsic fluctuations, probably originating by the poor precision of the nitrogen
determinations compared to other elements \citep{leisy06}.

The position of SMP~6 in the ON and CN planes can be interpreted 
either as a high metallicity PN with low mass progenitor, which eventually became carbon star under
the effects of TDU, or as the progeny of a low metallicity star of mass $\sim 3~\Msun$, whose
final surface chemistry reflects the effects of both TDU and HBB; the latter mechanism would
be responsible for the unusually large nitrogen observed. The large abundance of neon
(see Fig.~\ref{fnen}) would suggest the first hypothesis.

The nitrogen observed in SMP~26 is by far incompatible with low-mass stellar evolution,
rather pointing towards a higher mass progenitor. The overall CNO abundances suggest that
this object descends from a low metallicity star with mass at the edge of HBB ignition:
the carbon star stage was reached as a consequence of several TDU episodes, whereas the
high N was due to proton capture by some of the carbon nuclei transported to the 
surface.

\section{PNe in the SMC: a further test for AGB modelling}

The results presented here show that the majority of PNe in the SMC descend from low
mass progenitors, of initial mass below $2\Msun$. This conclusion holds for all PNe in 
group (a) and for a large 
fraction of the outliers, those in group (c). On this aspect the SMC is significantly
different from the LMC. In fact, as shown in paper I, the fraction of LMC
PNe descending from stars of mass above $\sim 3~\Msun$ is significantly larger than that of SMC PNe.
This result is however not surprising, if we consider the different star formation
histories (SFH) experienced by the two galaxies. The LMC SFH has two
prominent peaks at $\sim 100$ Myr and $\sim 500$ Myr \citep{harris09}, which favoured
the formation of stars of mass in the range $3-7~\Msun$; conversely, the SMC SFH
presents a single, extremely narrow peak at $\sim 500$ Myr \citep{harris04}, which is 
consistent with the progeny of the low-mass AGB stars observed.

The large majority of SMC PNe  are carbon rich, 
with a C/O ratio above unity; this indicates that all these objects descend from stars that 
reached the C-star stage during the AGB evolution. Only a small minority of SMC PNe are 
C-poor (and N-enhanced). This further difference between the SMC and LMC PNe
is explained by the average lower metallicity of SMC stars compared to those in the LMC.
Naturally, carbon stars(C/O$>$1) at low metallicity (and lower oxygen) are favoured since 
a smaller amount of carbon needs to be dredged up to reach the critical condition that 
define them. Furthermore, the depth of TDU increases with decreasing metallicity 
\citep{boothroyd88}, which enhances the surface carbon enrichment, thus the process goes 
in the same direction of favouring C-stars in the SMC with respect to the LMC.
The fact that the spectra of most of the SMC PNe observed presents the signature of
carbon-rich dust adds more robustness to this interpretation.

The most significant source of uncertainty in the modelling of low-mass AGB stars is the
extension of TDU, which drives the increase in the surface carbon. The efficiency of TDU
depends on physical parameters, primarily the core mass and the metallicity. These 
dependences are still highly uncertain, which reflects
our poor knowledge of how convection works in stars. Based on the comparison with the
luminosity function of carbon stars in the MC, synthetic AGB models have been extensively 
used to calculate the core mass at which TDU begins and the extension of the mixed region
\citep{martin93, marigo99, izzard04}. Detailed, standard models, where the convection/radiation
interface is fixed via the Schwartzschild criterion, could not reproduce the observations
\citep{karakas02}, thus suggesting that some overshoot from the base of the convective
envelope is required. The need for some extra-mixing was invoked by \citet{herwig07},
based on numerical simulations and, more recently, by \citet{kamath12}, on the basis of
the comparison between the observed and expected transition luminosity 
from O-rich to C-rich stars in clusters in the MC. Extra-mixing from the
bottom of the envelope has been used in almost the totality of the most recent 
works on AGB modelling \citep{cristallo09, weiss09, paperIV, garcia16}.
 
The carbon abundance of the PNe discussed here is $\log(C/H)+12<9$
(see right panel of Fig.~\ref{fpne}), which falls within the range of the 
evolutionary models; the C/O ratios never exceed $\sim 7$. 

The PNe belonging to group (b) exhibit a chemical composition with the clear imprinting of 
CNO nucleosynthesis. While common envelope evolution in binary systems or non standard
mixing of low-mass stars could partly explain this group of stars (see discussion at the
end of the previous section), it is likely that a few of these PNe have experienced HBB 
during the AGB evolution, which would explain the significant nitrogen enhancement and the 
low carbon and oxygen abundances measured. This would suggest massive 
progenitors, evolving on core masses above $\sim 0.9~\Msun$ \citep{ventura13}. For stars 
in this mass domain, convection is once more the major uncertainty, for
two reasons: a) the efficiency of convection determines the temperature at the base of the 
envelope \citep{vd05}, which is particularly relevant for the evolution of the surface oxygen;
b) the possible occurrence of TDU, still debated, would eventually increase the surface carbon, 
and, in some cases, produce high luminosity carbon stars \citep{frost98}. 

The surface chemical composition of group (b) PNe is nicely reproduced by our massive AGB
models of metallicity $Z=2\times 10^{-3}$. This result, if confirmed, 
would be the most reliable confirmation
obtained so far that low metallicity AGB stars of mass $M > 6~\Msun$ undergo a very 
advanced nucleosynthesis at the 
base of their envelope, as a results of HBB temperatures above $\sim 80$ MK. This finding
indicates that envelope convection is highly efficient in low-metallicity, massive 
AGB stars, as predicted by the FST schematisation adopted here. Note that a similar
result can be obtained within the traditional Mixing Length Theory description, provided
that a mixing length significantly higher than the standard vale used to reproduce the evolution of
the Sun is adopted \citep{doherty14}. Furthermore, the extremely low carbon abundances 
measured in these stars indicates that TDU has negligible (if any) effects on 
the evolution of the surface chemistry of these objects, at odds with other models 
for stars of similar mass and metallicity, which predict much higher quantities of carbon 
(see Fishlock et al. 2014 and references therein).

\section{Conclusions}
We characterise the PNe population of the SMC based on AGB models of various mass and
metallicity, which account for dust production in the circumstellar envelope.

We found that most of the PNe in the observed sample descend from stars with mass in the 
range $1~\Msun < M < 2~\Msun$ and metallicity $Z \sim 2-4\times 10^{-3}$, formed between
1~Gyr and 7~Gyr ago. Unlike their counterparts in the LMC, all SMC PNe have measured 
C/O ratio above unity; this is consistent with the evolutionary models, according to
which the achievement of the C-star stage is easier in stars of lower metallicity, as 
those currently evolving to PNe in the SMC. The analysis of the observed oxygen 
abundances, compared to the models, outlines that SMC PN progenitors formed out of a 
medium that was poorer in oxygen than the medium at the time of LMC PN progenitor 
formation, and that of the Milky Way. This is in agreement with the studies focused on 
the chemical evolution of the SMC.

The spread in carbon abundances observed in SMC PNe is nicely reproduced by the AGB 
models; this adds more robustness to the models used in the present investigation, 
particularly for what concerns the depth of the TDU experienced. 

Six PNe in the  SMC sample are greatly enhanced in nitrogen, thus indicating 
some CN (or CNO) activity. The chemical composition of these objects is reproduced by low 
metallicity stars of initial mass close to the threshold limit to undergo core collapse. 
Deep mixing during the red giant branch, common envelope evolution and binarity
are additional possibilities for this chemistry.
While for reasons related to the mass function and to the relatively short duration of 
the post-AGB phase of these objects it seems unlikely that all these stars descend from
massive AGB progenitors, it is reasonable to believe that part of these systems are the 
progeny of stars of mass in the range $6~\Msun < M < 8~\Msun$ and metallicity 
$Z \sim 4-8\times 10^{-3}$. 
The overall CNO chemistry of these PNe traces the equilibria of advanced HBB nucleosynthesis, 
corresponding to temperatures at the base of the convective envelope of the order
of $\sim 100$MK. If this hypothesis is confirmed, it would provide a strong indication
that in low-metallicity, massive AGB stars, convection is highly efficient. In 4 out of the
6 sources the carbon abundance is extremely small, thus indicating negligible effects of TDU.

We have now compared the best observed PN sample of the MCs to the AGB stellar 
evolutionary models of adequate metallicity. In the future we plan a similar study 
extended to Galactic PNe to increase the metallicity baseline of the comparison.

\section*{Acknowledgments}
MDC acknowledges the contribution of the FP7 SPACE project ASTRODEEP (Ref.No:312725), 
supported by the European Commission. DAGH was funded by the Ram\'on y Cajal fellowship 
number RYC$-$2013$-$14182 and he acknowledges support provided by the Spanish Ministry of 
Economy and Competitiveness (MINECO) under grant AYA$-$2014$-$58082-P.
FD acknowledges support from the Observatory of Rome.


\begin{thebibliography}{99}

\bibitem[Angelou et al.(2012)]{angelou12} Angelou G.~C., Stancliffe R.~J., 
Church R.~P., Lattanzio J.~C., Smith G.~H. \ 2012, ApJ, 749, 128
\bibitem[Bernard-Salas et al.(2009)]{2009ApJ...699.1541B} Bernard-Salas, J., 
Peeters, E., Sloan, G.~C., et al.\ 2009, ApJ, 699, 1541 
\bibitem[\protect\citeauthoryear{Bl\"ocker}{1995}]{blocker95}
Bl\"ocker T., 1995, A\&A, 297, 727
\bibitem[\protect\citeauthoryear{Bl\"ocker \& Sch\"oenberner}{1991}]{blocker91}
Bl\"ocker T., Sch\"oenberner D., 1991, A\&A, 244, L43
\bibitem[\protect\citeauthoryear{Boothroyd \& Sachmann}{1988}]{boothroyd88}
Boothroyd A.~I, Sachmann I.-J., 1988, ApJ, 328, 653
\bibitem[Boyer et al.(2011)]{boyer11} Boyer M.~L., Srinivasan S., 
van Loon, J.~T., et al.\ 2011, AJ, 142, 103 
\bibitem[\protect\citeauthoryear{Boyer et al.}{2012}]{boyer12}
Boyer M.~L., Srinivasan S., Riebel D., McDonald I., van Loon J.~Th., Clayton G.~C.,
Gordon K.~D., Meixner M., Sargent B.~A., Sloan G.~C., 2012, ApJ, 748, 40
\bibitem[\protect\citeauthoryear{Canuto \& Mazzitelli}{1991}]{cm91}
Canuto V.~M.~C., Mazzitelli I., 1991, ApJ, 370, 295
\bibitem[\protect\citeauthoryear{Cloutmann \& Eoll}{1976}]{cloutmann}
Cloutmann, L., \& Eoll, J.G.~1976, ApJ, 206, 548
\bibitem[Cristallo et al.(2009)]{cristallo09} Cristallo S., Straniero O., Gallino R., 
Piersanti L., Dominguez I., Lederer M.~T. \ 2009, ApJ, 696, 797
\bibitem[\protect\citeauthoryear{Dell'Agli et al.}{2014}]{flavia14}
Dell'Agli F., Ventura P., Garc{\'{\i}}a-Hern{\'a}ndez D.~A., Schneider R., Di Criscienzo M., 
Brocato E., D'Antona F., Rossi C., 2014, MNRAS, 442, L38
\bibitem[\protect\citeauthoryear{Dell'Agli et al.}{2015a}]{flavia15a}
Dell'Agli F., Ventura P., Schneider R., Di Criscienzo M., Garc{\'{\i}}a-Hern{\'a}ndez D.~A.,  
Rossi C., Brocato E.  2015a, MNRAS, 447, 2992
\bibitem[\protect\citeauthoryear{Dell'Agli et al.}{2015b}]{flavia15b}
Dell'Agli F., Garc{\'{\i}}a-Hern{\'a}ndez D.~A., Ventura P., Schneider R., Di Criscienzo M.,   
Rossi C. 2015b, MNRAS, 454, 4235
\bibitem[\protect\citeauthoryear{Di Criscienzo et al.}{2013}]{paperIII}
Di Criscienzo M., Dell'Agli F., Ventura P., Schneider R., Valiante R., 
La Franca F., Rossi C., Gallerani S., Maiolino, R., 2013, MNRAS, 433, 313
\bibitem[Doherty et al.(2014)]{doherty14} Doherty C.~L., Gil-Pons P., Lau H.~H.~B., 
Lattanzio J.~C., Siess L.\ 2014, MNRAS, 437, 195 
\bibitem[\protect\citeauthoryear{Ferrarotti \& Gail}{2006}]{fg06}
Ferrarotti A.~D., Gail H.~P., 2006, A\&A, 553, 576
\bibitem[Jacoby(1980)]{1980ApJS...42....1J} Jacoby, G.~H.\ 1980, ApJS, 42, 1 
\bibitem[Fishlock et al.(2014)]{fishlock14} Fishlock C.~K., 
Karakas A.~I., Lugaro M., Yong, D.\ 2014, ApJ, 797, 44 
\bibitem[Frost et al.(1998)]{frost98} Frost C.~A., Cannon R.~C., Lattanzio J.~C., Wood P.~R., 
Forestini M.\ 1998, A\&A, 332, L17 
\bibitem[Garc{\'{\i}}a-Hern{\'a}ndez et al.(2006)]{garcia06} 
Garc{\'{\i}}a-Hern{\'a}ndez D.~A., Garc{\'{\i}}a-Lario P., Plez B., et 
al.\ 2006, Science, 314, 1751 
\bibitem[Garc{\'{\i}}a-Hern{\'a}ndez et al.(2007)]{garcia07} 
Garc{\'{\i}}a-Hern{\'a}ndez D.~A., Garc{\'{\i}}a-Lario P., Plez B., 
et al.\ 2007, A\&A, 462, 711 
\bibitem[Garc{\'{\i}}a-Hern{\'a}ndez et al.(2009)]{garcia09} 
Garc{\'{\i}}a-Hern{\'a}ndez D.~A., Manchado A., Lambert D.~L., et al.\ 
2009, ApJL, 705, L31 
\bibitem[Garc{\'{\i}}a-Hern{\'a}ndez et al.(2016)]{garcia16} 
Garc{\'{\i}}a-Hern{\'a}ndez,D.~A., Ventura P., Delgado-Inglada G., Dell'Agli F.,
Di Criscienzo M., Yague A.\ 2016, MNRAS 
\bibitem[Gilroy(1989)]{gilroy89} Gilroy K.~K.\ 1989, ApJ, 347, 835 
\bibitem[Gilroy \& Brown(1991)]{gilroy91} Gilroy K.~K., Brown J.~A.\ 1991, ApJ, 371, 578 
\bibitem[Girardi \& Marigo(2007)]{girardi07} Girardi L., Marigo P.\ 2007, A\&A, 462, 237
\bibitem[\protect\citeauthoryear{Grevesse \& Sauval}{1998}]{gs98}
Grevesse N., Sauval A.~J / 1998, SSrv, 85, 161
\bibitem[Groenewegen \& de Jong(1993)]{martin93} Groenewegen M.~A.~T., \& 
de Jong, T.\ 1993, A\&A, 267, 410 
\bibitem[Groenewegen et al.(2007)]{martin07} 
Groenewegen M.~A.~T., Wood P.~R., Sloan G.~C., et al.\ 2007, MNRAS, 376, 313 
\bibitem[\protect\citeauthoryear{Harris \& Zaritsky}{2004}]{harris04}
Harris J. \& Zaritsky D.\ 2004, AJ, 127, 1531
\bibitem[\protect\citeauthoryear{Harris \& Zaritsky}{2009}]{harris09}
Harris J. \& Zaritsky D. 2009, ApJ, 138, 1243
\bibitem[Herwig(2005)]{herwig05} Herwig F.\ 2005, ARA\&A, 43, 435 
\bibitem[Herwig et al.(2007)]{herwig07} Herwig F., Freytag B., Fuchs T., et al.\ 2007, 
Why Galaxies Care About AGB Stars: Their Importance as Actors and Probes, 378, 43 
\bibitem[Izzard et al.(2004)]{izzard04} Izzard R.~G., Tout C.~A., Karakas A.~I.,
Pols, O.~R.\ 2004, MNRAS, 350, 407 
\bibitem[Kamath et al.(2012)]{kamath12} Kamath D., Karakas A.~I., Wood, P.~R.\ 2012, ApJ, 746, 20 
\bibitem[Karakas et al.(2002)]{karakas02} Karakas A.~I., Lattanzio J.~C., Pols O.~R.\ 2002, 
PASA, 19, 515 
\bibitem[Karakas \& Lattanzio(2014)]{karakas14} Karakas A.~I., Lattanzio J.~C.\ 2014, PASA, 31, e030 
\bibitem[Keller \& Wood(2006)]{keller06} Keller S.~C., Wood P.~R.\ 2006, ApJ, 642, 834 
\bibitem[Leisy \& Dennefeld(2006)]{leisy06} Leisy P., Dennefeld, M.\ 2006, A\&A, 456, 451 
\bibitem[Leisy \& Dennefeld(1996)]{leisy96} Leisy P., Dennefeld, M.\ 1996, A\&AS, 116, 95 
\bibitem[Maraston et al.(2006)]{maraston06} Maraston C., Daddi E., Renzini A., et al.\ 2006, 
ApJ, 652, 85 
\bibitem[Marigo et al.(1999)]{marigo99} Marigo P., Girardi L., Bressan, A.\ 1999, A\&A, 344, 123 
\bibitem[Marigo et al.(2003)]{marigo03} Marigo P., Bernard-Salas J., Pottasch S.~R., 
Tielens A.~G.~G.~M., Wesselius P.~R.\ 2003, A\&A, 409, 619 
\bibitem[\protect\citeauthoryear{Marigo \& Aringer}{2009}]{marigo09} 
Marigo P., Aringer B., 2009, A\&A, 508, 1538
\bibitem[Marigo et al.(2011)]{marigo11} Marigo P., Bressan A., Girardi L., 
et al.\ 2011, Why Galaxies Care about AGB Stars II: Shining 
Examples and Common Inhabitants, 445, 431 
\bibitem[Meyssonnier \& Azzopardi(1993)]{1993A&AS..102..451M} Meyssonnier, N., \& Azzopardi, M.\ 1993, A\&AS 102, 451 
\bibitem[Morgan \& Good(1985)]{1985MNRAS.213..491M} Morgan, D.~H., \& Good, A.~R.\ 1985, MNRAS 213, 491 
\bibitem[\protect\citeauthoryear{Nanni et al.}{2013a}]{nanni13a} 
Nanni A., Bressan A., Marigo P., Girardi L., 2013a, MNRAS, 434, 488
\bibitem[\protect\citeauthoryear{Nanni et al.}{2013b}]{nanni13b} 
Nanni A., Bressan A., Marigo P., Girardi L., 2013b, MNRAS, 434, 2390
\bibitem[\protect\citeauthoryear{Nanni et al.}{2014}]{nanni14} 
Nanni A. Bressan A. Marigo P. Girardi L., 2014, MNRAS, 438, 2328
\bibitem[Porter et al.(2005)]{2005ApJ...622L..73P} Porter, R.~L., Bauman, 
R.~P., Ferland, G.~J., \& MacAdam, K.~B.\ 2005, ApJL, 622, L73 
\bibitem[Pottasch \& Bernard-Salas (2006)]{pottasch06} Pottasch S.~R., Bernard- Salas J. 
\ 2006, A\&A, 457, 189
\bibitem[\protect\citeauthoryear{Renzini \& Voli}{1981}]{renzini81} Renzini A.,
Voli M., 1981, A\&A, 94, 175
\bibitem[\protect\citeauthoryear{Riebel et al.}{2010}]{riebel10} 
Riebel D., Meixner M., Fraser O., Srinivasan S., Cook K., Vijh U., 2010, ApJ, 723, 1195
\bibitem[\protect\citeauthoryear{Riebel et al.}{2012}]{riebel12} 
Riebel D., Srinivasan S., Sargent B., Meixner M., 2012, AJ, 753, 71
\bibitem[Romano et al.(2010)]{romano10} Romano D., Karakas A.~I., Tosi M., 
Matteucci, F.\ 2010, A\&A, 522, A32
\bibitem[Russell \& Dopita (1992)]{russell92} Russell S.~C., Dopita M.~A. \ 1992, ApJ, 384, 508
\bibitem[Sanduleak \& Pesch(1981)]{1981PASP...93..431S} Sanduleak, N., \& Pesch, P.\ 1981, PASP, 93, 431 
\bibitem[Sanduleak et al.(1978)]{1978PASP...90..621S} Sanduleak, N., MacConnell, D.~J., \& Philip, A.~G.~D.\ 1978, PASP,  90, 621 
\bibitem[Santini et al.(2014)]{santini14} Santini P., Maiolino R., Magnelli B., 
et al.\ 2014, A\&A, 562, A30
\bibitem[\protect\citeauthoryear{Spite et al.}{1989}]{spite89} 
Spite M., Barbuy B., Spite F. \ 1989, A\&A, 222, 35
\bibitem[\protect\citeauthoryear{Srinivasan et al.}{2009}]{srinivasan09} 
Srinivasan S. et al., 2009, AJ, 137, 4810
\bibitem[Srinivasan et al.(2011)]{srinivasan11} Srinivasan S., Sargent B.~A., 
Meixner, M.\ 2011, A\&A, 532, A54 
\bibitem[Shaw et al.(2010)]{2010ApJ...717..562S} Shaw, R.~A., Lee, T.-H., Stanghellini, L., et al.\ 2010, ApJ, 717, 562 
\bibitem[Stanghellini et al.(1999)]{1999ApJ...510..687S} Stanghellini, L., Blades, J.~C., Osmer, S.~J., Barlow, M.~J., \& Liu, X.-W.\ 1999, ApJ, 510, 687 
\bibitem[Stanghellini et al.(2000)]{letizia00} Stanghellini L., 
Shaw R.~A., Balick B., Blades J.~C.\ 2000, ApJL, 534, L167 
\bibitem[Stanghellini et al.(2003)]{Stanghelini2003a} Stanghellini, L., 
Villaver, E., Shaw, R.~A., \& Mutchler, M.\ 2003, ApJ, 598, 1000 
\bibitem[Stanghellini et al.(2003)]{Stanghellini2003b} Stanghellini, L., Shaw, R.~A., Balick, B., et al.\ 2003, ApJ, 596, 997 
\bibitem[Stanghellini et al.(2005)]{letizia05} Stanghellini, L., 
Shaw, R.~A., \& Gilmore, D.\ 2005, ApJ, 622, 294 
\bibitem[Stanghellini et al.(2007)]{letizia07} Stanghellini L., 
Garcia-Lario P., Garc{\'{\i}}a-Hern{\'a}ndez D.~A., et al., 2007, ApJ, 671, 1669
\bibitem[Stanghellini et al.(2009)]{letizia09} Stanghellini L., 
Lee T.-H., Shaw R.~A., Balick B., Villaver E.\ 2009, ApJ, 702, 733 
\bibitem[Stasi{\'n}ska et al.(1998)]{1998A&A...336..667S} Stasi{\'n}ska, G., Richer, M.~G., \& McCall, M.~L.\ 1998, A\&A, 336, 667 
\bibitem[Valiante et al.(2011)]{valiante11} Valiante R., 
Schneider R., Salvadori S., Bianchi S.\ 2011, MNRAS, 416, 1916
\bibitem[Ventura et al.(2000)]{ventura00} Ventura P., D'Antona F., Mazzitelli I.\ 2000, A\&A, 363, 605 
\bibitem[Ventura et al.(2001)]{ventura01} Ventura P., D'Antona F., Mazzitelli I., 
Gratton R.\ 2001, ApJL, 550, L65 
\bibitem[\protect\citeauthoryear{Ventura \& D'Antona}{2005}]{vd05}
Ventura P., D'Antona F., 2005, A\&A, 431, 279
\bibitem[\protect\citeauthoryear{Ventura \& D'Antona}{2009}]{ventura09} 
Ventura P., D'Antona F., 2009, MNRAS, 499, 835
\bibitem[\protect\citeauthoryear{Ventura et al.}{2012a}]{paperI} 
Ventura P., Di Criscienzo M., Schneider R., Carini R., Valiante R., D'Antona F., 
Gallerani S., Maiolino R., Tornamb\'e A., 2012a, MNRAS, 420, 1442
\bibitem[\protect\citeauthoryear{Ventura et al.}{2012b}]{paperII} 
Ventura P., Di Criscienzo M., Schneider R., Carini R., Valiante R., D'Antona F., 
Gallerani S., Maiolino R., Tornamb\'e A., 2012b, MNRAS, 424, 2345
\bibitem[\protect\citeauthoryear{Ventura et al.}{2014a}]{paperIV} 
Ventura P., Dell'Agli F., Di Criscienzo M., Schneider R., Rossi C., La Franca F., 
Gallerani S., Valiante R., 2014a, MNRAS, 439, 977
\bibitem[Ventura et al.(2014b)]{ventura14b} Ventura, P., 
Di Criscienzo, M.~D., D'Antona, F., et al.\ 2014, MNRAS, 437, 3274 
\bibitem[\protect\citeauthoryear{Ventura et al.}{2013}]{ventura13} 
Ventura P., Di Criscienzo M., Carini R., D'Antona F., 2013, MNRAS, 431, 3642
\bibitem[\protect\citeauthoryear{Ventura et al.}{1998}]{ventura98} Ventura P.,
Zeppieri A., Mazzitelli I., D'Antona F., 1998, A\&A, 334, 953
\bibitem[Ventura et al.(2015)]{ventura15} Ventura P., Karakas A.~I., Dell'Agli F., 
Boyer M.~L., Garc{\'{\i}}a-Hern{\'a}ndez D.~A., Di Criscienzo M., Schneider R. \ 2015, MNRAS, 450, 3181 
\bibitem[Ventura et al.(2016)]{ventura16} Ventura P., Karakas A.~I., Dell'Agli F., 
Garc{\'{\i}}a-Hern{\'a}ndez D.~A., Boyer M.~L., Di Criscienzo M. \ 2016, arXiv:1601.01845 
\bibitem[Villaver et al.(2004)]{2004ApJ...614..716V} Villaver, E., Stanghellini, L., \& Shaw, R.~A.\ 2004, ApJ, 614, 716 
\bibitem[\protect\citeauthoryear{Wachter et al.}{2002}]{wachter02} 
Wachter A., Schr\"oder K.~P., Winters J.~M., Arndt T.~U., Sedlmayr E., 2002, A\&A, 384, 452
\bibitem[\protect\citeauthoryear{Wachter et al.}{2008}]{wachter08} 
Wachter A., Winters J.~M., Schr\"oder K.~P., Sedlmayr E., 2008, A\&A, 486, 497
\bibitem[Weiss \& Ferguson(2009)]{weiss09} Weiss A., Ferguson, J.~W.\ 2009, A\&A, 508, 1343 
\end{thebibliography}
\end{document}